\documentclass[aps,prl,twocolumn,groupedaddress,floatfix,sort&compress,longnamesfirst]{revtex4-2}

\usepackage{amsmath,amssymb}
\usepackage{bm,bbm}
\usepackage{hyperref}
\hypersetup{
	colorlinks=true,
	linkcolor=blue,
	filecolor=blue,
	urlcolor=blue,
	citecolor=blue,
}
\usepackage{graphicx}
\graphicspath{{figure/}}

\begin{document}


\title{Time Crystal in a Single-mode Nonlinear Cavity}


\author{Yaohua Li$^{1}$}
\author{Chenyang Wang$^{1}$}
\author{Yuanjiang Tang$^{1}$}
\author{Yong-Chun Liu$^{1, 2,}$}
\email{ycliu@tsinghua.edu.cn}
\affiliation{$^{1}$State Key Laboratory of Low-Dimensional Quantum Physics, Department of Physics, Tsinghua University, Beijing 100084, P. R. China}
\affiliation{$^{2}$Frontier
Science Center for Quantum Information, Beijing 100084, China}


\date{\today}

\begin{abstract}
	Time crystal is a class of non-equilibrium phases with broken time-translational symmetry. Here we demonstrate the time crystal in a single-mode nonlinear cavity. The time crystal originates from the self-oscillation induced by a linear gain and is stabilized by a nonlinear damping. We show in the time crystal phase there are sharp dissipative gap closing and pure imaginary eigenvalues of the Liouvillian spectrum in the thermodynamic limit. Dynamically, we observe a metastable regime with the emergence of quantum oscillation, followed by a dissipative evolution with a time scale much smaller than the oscillating period. Moreover, we show there is a dissipative phase transition at the Hopf bifurcation of the model, which can be characterized by the photon number fluctuation in the steady state. These results pave a new promising way for further experiments and deepen our understanding of time crystals.
\end{abstract}
\maketitle

{\it Introduction.}---The time crystal is an analogy of the spacial crystal, and it is defined as a phase of matter where the time-translational symmetry of the ground state is spontaneously broken. The concept of time crystal was first proposed by Shapere and Wilczek \cite{PhysRevLett.109.160402,PhysRevLett.109.160401,PhysRevLett.109.163001,PhysRevLett.111.250402}. However, the initially proposed model with spontaneously rotating ground state is ruled out by no-go theorems in subsequent studies \cite{PhysRevLett.110.118901,PhysRevLett.111.070402,PhysRevLett.114.251603}. The no-go theorems still allow closed systems with long-range couplings to break the time-translational symmetry, as verified in a spin-$1/2$ many-body Hamiltonian \cite{PhysRevLett.123.210602}. However, this model is also in dispute \cite{khemani_comment_2020,kozin_reply_2020}. The concept of time crystal is also generalized into periodically driven systems, as known as the discrete time crystal \cite{RevModPhys.95.031001,PhysRevA.91.033617,PhysRevLett.117.090402,PhysRevLett.118.030401,PhysRevLett.120.110603,PhysRevLett.123.150601,yao_classical_2020,PhysRevLett.130.120403}. It is characterized by subharmonic oscillations and thus breaks the discrete time-translational symmetry. The discrete time crystal has already been realized in trapped ions \cite{zhang_observation_2017,kyprianidis_observation_2021}, nuclear spins \cite{choi_observation_2017,PhysRevLett.120.180603,PhysRevLett.120.180602,osullivan_signatures_2020,randall_many-bodylocalized_2021} and other systems \cite{PhysRevLett.121.185301,PhysRevLett.120.215301,PhysRevA.105.012418,mi_time-crystalline_2022,zhang_digital_2022,doi:10.1126/sciadv.abm7652}.

Introducing couplings to the environment is another method to circumvent the no-go theorem and to obtain time crystal dynamics \cite{PhysRevLett.120.040404,PhysRevLett.127.043602,PhysRevLett.127.253601,taheri_all-optical_2022,PhysRevLett.131.056502}. This method can still start from a time-independent Hamiltonian and break the continuous time-translational symmetry \cite{buca_non-stationary_2019,PhysRevLett.130.150401,PhysRevLett.130.180401,buca_dissipation_2019,zhu_dicke_2019,paulino_nonequilibrium_2023}. Specifically, nonlinear systems with limit cycles are shown to be powerful platforms to realize this kind of time crystal \cite{PhysRevA.99.053605,kongkhambut_observation_2022,liu_photonic_2023,kosior_nonequilibrium_2023,kosior_nonequilibrium_2023}. In classical systems, two essential conditions are proposed to distinguish a time crystal from normal limit cycle models: the oscillating phase takes random values from 0 to $2\pi$ for repeated realizations; the oscillation dynamics is robust against perturbations or fluctuations \cite{kongkhambut_observation_2022}. In quantum systems, the time crystal can be directly characterized by pure imaginary eigenvalues in the Liouvillian spectrum in the thermodynamic limit \cite{PhysRevLett.121.035301,lledo_driven_2019,PhysRevA.101.033839,bakker_driven-dissipative_2022,PhysRevA.107.L010201}. However, it is still difficult to obtain a full quantum description of the time crystal in open and nonlinear systems, as these systems contain both many-body interactions and couplings to the environment.

\begin{figure}[b]
	\centering
	\includegraphics[width=0.4\textwidth]{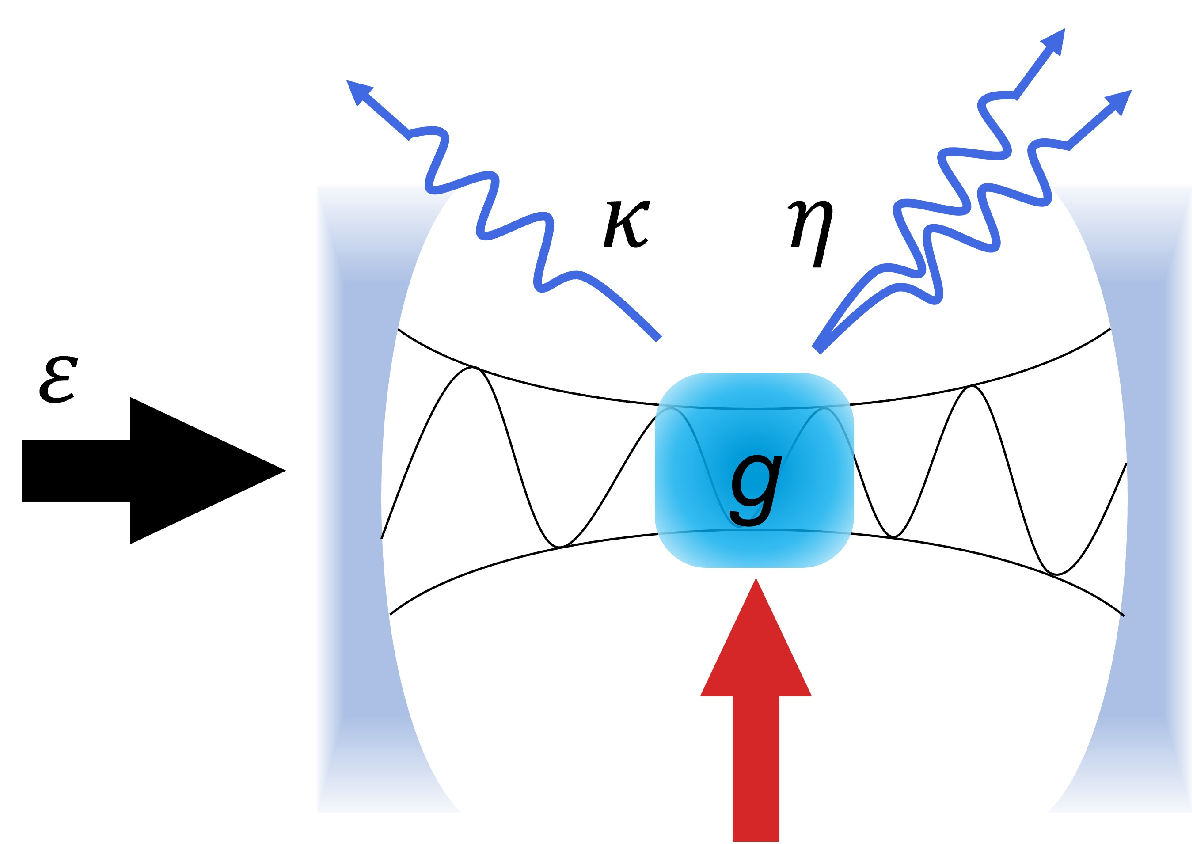}
	\caption{Schematic representation of our setup: a cavity mode $a$ under driving with strength $\varepsilon$. There are both linear (single-photon) and nonlinear (two-photon) damping with rates $\kappa$ and $\eta$, respectively. There is also a linear gain with rate $g$.}
	\label{fig:a}
\end{figure}

Here we demonstrate that the time crystal can be realized in a single-mode nonlinear cavity. This model can be exactly solved for large Fock space dimensions. Surprisingly, we show such a simple model is enough to reveal the rich properties of a time crystal. We observe a sharp dissipative gap closing and the emergence of pure imaginary eigenvalues of Liouvillian spectrum in the thermodynamic limit. We show the system will quickly enter a metastable regime with quantum oscillation. Afterwards there is a dissipative evolution towards the steady state with a time scale much smaller than the oscillating period. In this process, the phase fluctuation gradually smears out the quantum limit cycle, and finally the quantum oscillation disappears. Differently, we find that the oscillating phases of classical phase trajectories are encoded in the initial distribution without a dephasing preocess. Moreover, we show the classical Hopf bifurcation corresponds to a dissipative phase transition. It is verified by both the dissipative gap closing and the sharp change of steady-state properties. We find the photon number fluctuation is a good order parameter to characterize the dissipative phase transition.

{\it Hopf bifurcation and classical limit cycle.}---We consider a single-mode cavity described by the master equation ($\hbar = 1$):
\begin{equation}\label{eq:master}
	\dot{\rho} = i[\rho, H] + \kappa \mathcal{D}[a]\rho + g \mathcal{D}[a^{\dag}]\rho + \eta \mathcal{D}[a^{2}]\rho,
\end{equation}
where $\rho$ is the density matrix, $a$ ($a^{\dag}$) is the bosonic annihilation (creation) operator, and $\mathcal{D}(o)\rho=o\rho o^{\dag}-(o^{\dag}o\rho+\rho o^{\dag}o)/2$ is the Liouvillian for operator $o$. We have standard terms for linear (single-photon) damping, linear gain, and nonlinear (two-photon) damping \cite{leghtas_confining_2015} with rates $\kappa$, $g$ and $\eta$, respectively [see Fig. \ref{fig:a}]. The Hamiltonian with driving is $H = -\Delta a^{\dag}a + \varepsilon a^{\dag} + \varepsilon^{*}a$, where $\Delta$ is the detuning between the driving frequency and the cavity frequency, and $\varepsilon$ is the driving strength.

\begin{figure}[t]
	\centering
	\includegraphics[width=0.48\textwidth]{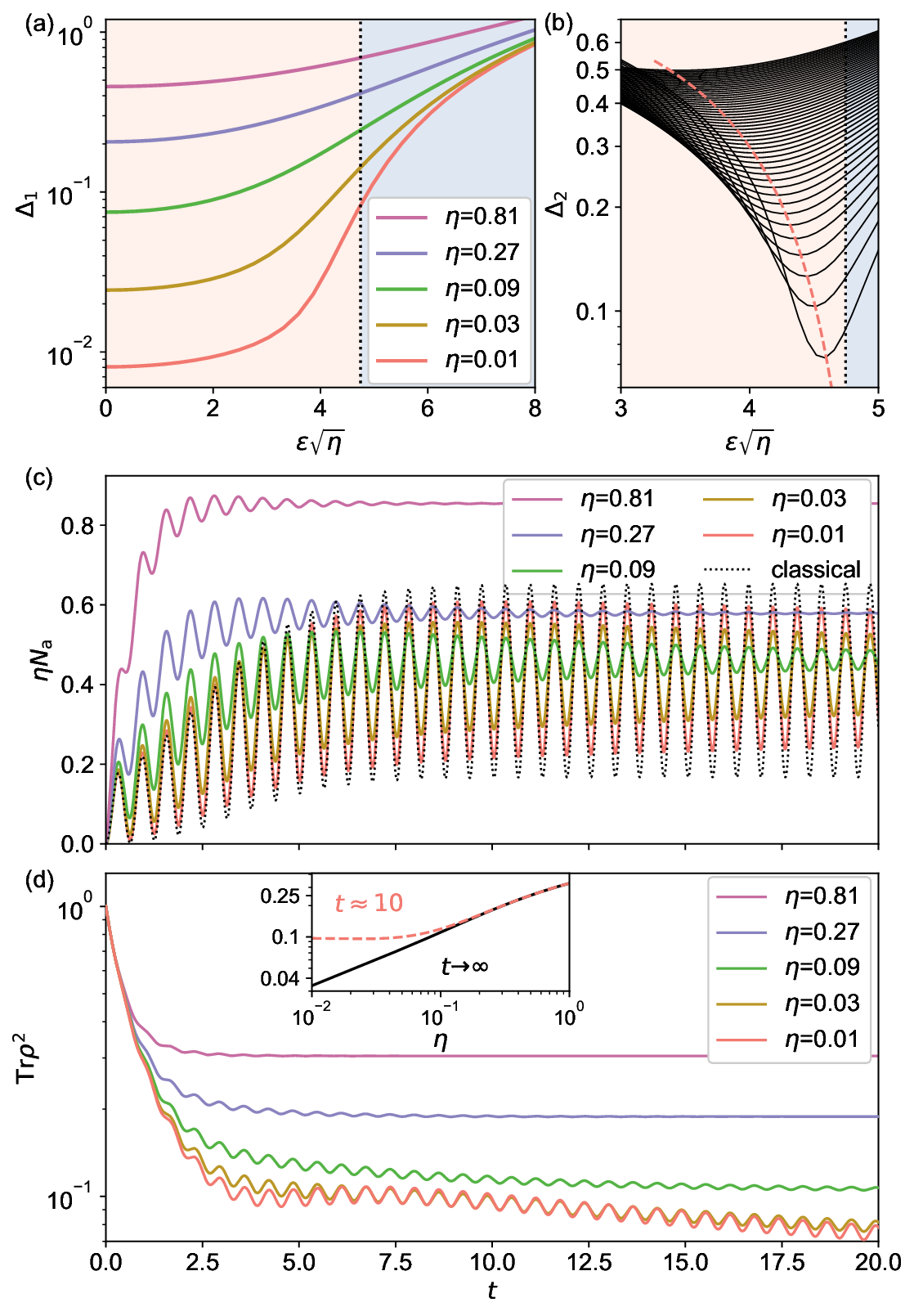}
	\caption{Dissipative gap closing and stable quantum oscillation. (a),(b) Two dissipative gaps as a function of rescaled driving strength $\varepsilon\sqrt{\eta}$ for different nonlinear damping rate $\eta$. The black dashed line indicates the Hopf bifurcation of the reduced classical model, and the pink (blue) area indicates the limit cycle (equilibrium) phase. The fifty lines in (b) from the bottom up correspond to equally spaced $\eta$ from 0.001 to 0.05. The dashed red line in (b) indicates the nonlinear fitting curve of the minimums. (c),(d) The time evolution of rescaled photon number $\eta N_{\mathrm{a}}$ (c) and state purity $\mathrm{Tr}\rho^{2}$ (d). The black dashed line in (c) indicates the time evolution of the classical model. The inset in (d) shows the average purity in a period around $t=10$ (red dashed line) and the steady-state purity for $t\to\infty$ (black solid line) as a function of the nonlinear damping rate. The initial state is the vacuum state, and the rescaled driving strength is chosen as $\varepsilon\sqrt{\eta}=2$. Other parameters are $\Delta = 10$, $\kappa = 0.1$, and $g = 1$.}
	\label{fig:b}
\end{figure}

This model can be reduced to a driven Van der Pol (VdP) oscillator under the mean-field approximation \cite{weiss_quantum-coherent_2017,navarrete-benlloch_general_2017,sonar_squeezing_2018,ben_arosh_quantum_2021}, where the evolution of amplitude $\alpha = \langle a\rangle$ is appropriately governed by
\begin{equation}\label{eq:classical}
	\dot{\alpha} = (-\frac{\kappa-g}{2}+i\Delta)\alpha -\eta |\alpha|^{2}\alpha -i\varepsilon.
\end{equation}
It is a famous model that has Hopf bifurcation and limit cycle phase. The Hopf bifurcation point, characterized by the rescaled driving strength $\varepsilon\sqrt{\eta}$, is a constant. This constant can be approximately obtained by letting $\eta\to0$, which is
\begin{equation}
	\varepsilon\sqrt{\eta}\approx \frac{\sqrt{(g-\kappa)[(g-\kappa)^{2}+4\Delta^{2}]}}{4}.
\end{equation}
The dashed black line in Fig. \ref{fig:b}(a) represents the Hopf bifurcation, and the pink (blue) area denotes the limit cycle (equilibrium) phase of the reduced classical model.

{\it Dissipative gap closing and quantum oscillation.}---Beyond the well-established classical model, we are more interested in the non-equilibrium behavior at the full quantum level. We numerically solve the Liouvillian spectrum and the time evolution of the cavity mode in a truncated Fock basis \cite{sup}. From the Liouvillian spectrum, we compute two kinds of ``dissipative gaps" $\Delta_{1,2}$, which are the largest real part of the eigenvalues with a nonzero and zero imaginary part, respectively.

In Fig. \ref{fig:b}(a) and \ref{fig:b}(b), we show the dissipative gaps as a function of the rescaled driving strength $\varepsilon\sqrt{\eta}$ for different nonlinear damping rate $\eta$. There is a rapid closing of the first dissipative gap $\Delta_{1}$ in the limit cycle phase when the nonlinear damping rate $\eta$ approaches zero, i.e., the thermodynamic limit or the weak interaction limit. Similar results have also been obtained in two-mode cavities \cite{bakker_driven-dissipative_2022}. However, due to the much larger dimensions of truncated Hilbert space that we can use, we obtain a much smoother dissipative gap-closing pattern, where the phase boundary matches well with the Hopf bifurcation. The closing of the second dissipative gap $\Delta_{2}$ also indicates a kind of dissipative phase transition \cite{PhysRevA.86.012116,PhysRevA.97.013825} near the Hopf bifurcation.

Importantly, the closing dissipative gap indicates a long-time quantum oscillation and thus indicates the emergence of the time crystal. The time crystal originates from the self-oscillation induced by the linear gain and is stabilized by the nonlinear damping, similar to the mechanism of the limit cycle emerging in the classical limit. At the quantum level, the real part of the Liouvillian spectrum is always non-positive, and there is always a steady state. However, the time arriving at the steady state becomes infinite when the dissipative gap closes in the thermodynamic limit. Consequently, we also denote the limit cycle phase as the time crystal phase in the following.

To capture the non-equilibrium oscillating behavior at the quantum level, we compute the time evolution of photon number $N_{\mathrm{a}}=\langle a^{\dag}a\rangle$ for five different values of nonlinear damping rate $\eta$, as shown in Fig. \ref{fig:b}(c). The initial state is a vacuum state, and the rescaled driving strength is all chosen as $\varepsilon\sqrt{\eta}=2$. The vertical value is rescaled photon number (the photon number multiplied by the nonlinear damping rate $\eta N_{\mathrm{a}}$) to eliminate the amplitude amplification for small nonlinear damping rates. The rescaling of variables are a standard method to find the quantum expectation values approaching the classical limit. We also plot the classical time evolution governed by Eq. (\ref{eq:classical}) starting from $\alpha=0$ (black dashed line). As we reduce the nonlinear damping rate, a stable quantum oscillation emerges with both increasing oscillation amplitude and increasing relaxation time. This oscillating behavior verifies the closing dissipative gap in the thermodynamic limit and thus proves itself a time crystal. When the nonlinear damping rate approaches zero, the time evolution at the quantum level approaches the classical time evolution.

The quantum-classical transition is more clear when we look at the state purity $\mathrm{Tr}\rho^{2}$. The steady-state state purity approaches zero with a fixed scaling when we decrease the nonlinear damping rate, as shown in the inset (black solid lines) of Fig. \ref{fig:b}(d). For smaller nonlinear damping rates, the driving strength is enlarged as $\varepsilon\sqrt{\eta}$ remains fixed. In this case, the steady state is driven into a higher-occupation state in the Fock space and more Fock basises are included. Consequently, the steady state is more like a classical equilibrium state in the thermodynamic limit.
In Fig. \ref{fig:b}(d), we also plot the time evolution of state purity, which contains rich phenomena that have not been found in other time crystal models before. The initial state is the vacuum state with purity $\mathrm{Tr}\rho^{2}=1$. In the beginning, the state purity decreases quickly with a similar rate for different nonlinear damping rates. After the quick decrease, the system enters an oscillating regime before the final relaxation to the steady state. It is a metastable regime, originating from a large separation between two consecutive eigenvalues \cite{macieszczak_towards_2016,labay-mora_quantum_2023}. Moreover, the state purity is partially preserved in the metastable regime. It is more clear when we analyze the scaling behaviour of the purity near $t=10$. As indicated by the red dashed line in the inset of Fig. \ref{fig:b}(d), the state purity near $t=10$ does not decrease when we further reduce the nonlinear damping rate, which is distinct from the behavior of the steady-state purity. It reveals that this time crystal phase can partially preserve the information of the initial state in the metastable regime and may find application in quantum associative memory \cite{labay-mora_quantum_2023,sup}.

\begin{figure}[t]
	\centering
	\includegraphics[width=0.48\textwidth]{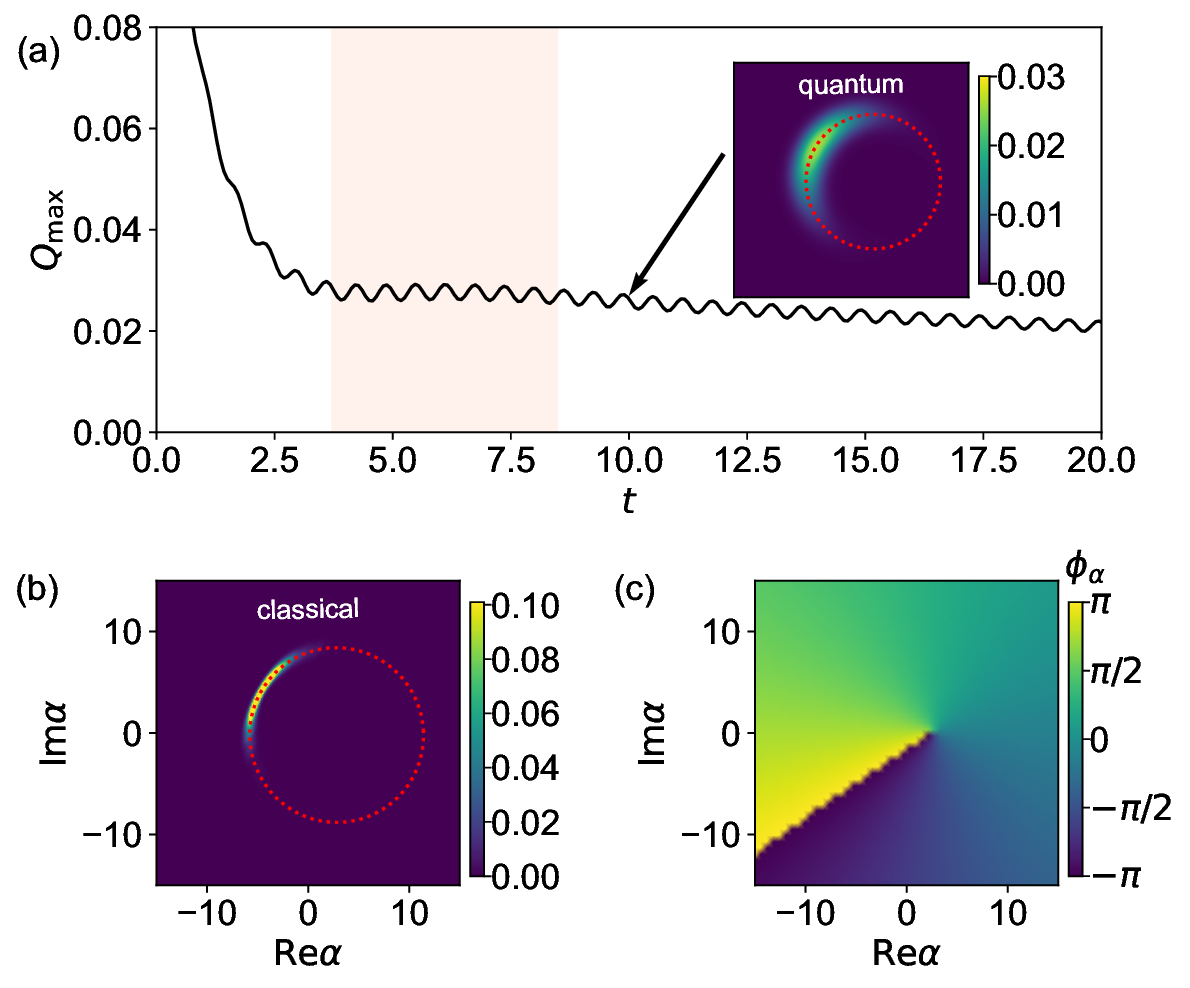}
	\caption{Dephasing in the quantum and classical evolution. (a) Maximal values of the quantum Husimi distribution as a function of time. The pink area denotes the metastable regime. The inset is the detailed quantum Husimi distribution at $t=10$. (b) Classical Husimi distribution at $t = 10$. The initial states are both the vacuum state. (c) Oscillating phases in the classical trajectory at $t = 10$. The oscillating phase depends on the start point in the phase space. There is a singular point where the oscillating phase is not continuous. The parameters are $\Delta = 10$, $\kappa = 0.1$, $g = 1$, $\eta = 0.005$, and $\varepsilon\sqrt{\eta} = 2$.}
	\label{fig:c}
\end{figure}

{\it Husimi Q-function.}---Next, we investigate the quantum oscillation from a quasiprobability distribution, i.e., the Husimi Q-function defined as $Q_{\mathrm{q}}(\alpha)=\langle \alpha |\rho|\alpha\rangle$. Figure \ref{fig:c}(a) shows the maximal values of the Husimi distribution in the time evolution starting from an initial vacuum state \cite{qutip}. The maximal value can partially reflect the phase fluctuation of the quantum oscillation. Similar to the evolution of the state purity, there is a fast dephasing process at the beginning and a slow dephasing process at the end \cite{sup}. In between, there is a metastable regime where the phase fluctuations remain unchanged. Finally, the phase fluctuation will smear out the quantum limit cycle and the quantum oscillation will disappear.

As a comparison, we also plot in Fig. \ref{fig:c}(b) the classical probability distribution at $t = 10$. It is obtained through the Monte Carlo simulations with 10000 trajectories, where the initial distribution also uses the vacuum state. In the reduced classical model the probability distribution directly decays to a part of the limit cycle without dephasing. The red dashed lines in Fig. \ref{fig:c}(a) and \ref{fig:c}(b) denote the limit cycle.

\begin{figure*}[t]
	\centering
	\includegraphics[width=0.9\textwidth]{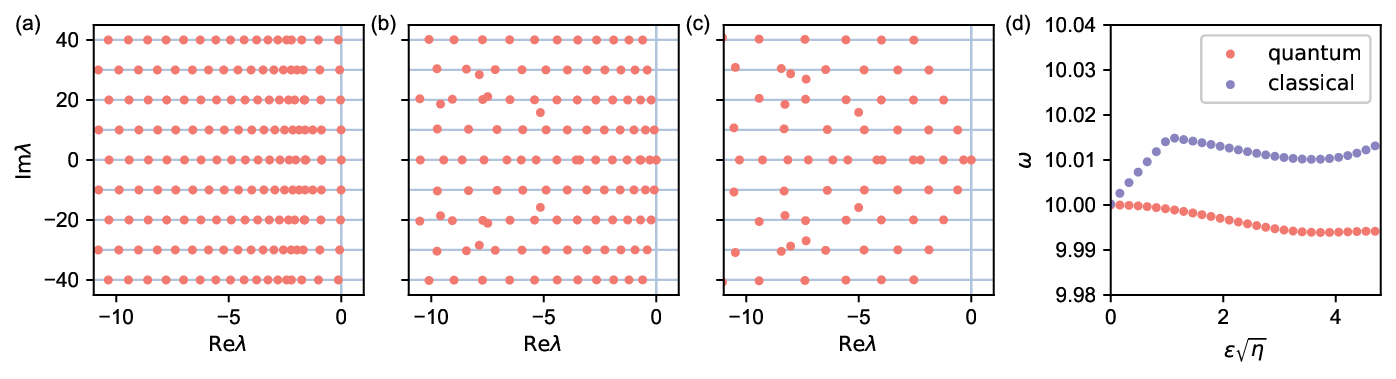}
	\caption{Liouvillian spectra for $\varepsilon\sqrt{\eta}=0$ (a), $\varepsilon\sqrt{\eta}=4.8$ (b) and $\varepsilon\sqrt{\eta}=7.2$ (c), corresponding to time crystal phase, critical regime and equilibrium phase. The red points are the eigenvalues, while the horizontal blue lines denote a frequency scale equal to the detuning $\Delta$. (d) The oscillating frequencies as a function of the rescaled driving strength. The quantum results (red points) are the imaginary part of the eigenvalues with the largest real part and nonzero imaginary part. The classical results (blue points) are obtained by solving the limit cycle through the Poincare map. Other parameters are $\Delta = 10$, $\kappa = 0.1$, $g = 1$, and $\eta = 0.01$.}
	\label{fig:d}
\end{figure*}

We further show in Fig. \ref{fig:c}(c) the oscillating phases at $t=10$ versus the place of the initial state in the phase space. The oscillating phase is continuous when the initial state changes in the phase space except for a singular point. Starting from this point, the oscillating phase is highly uncertain as an infinitely small error in the initial state can lead to any possible phase in the final limit cycle. The probability distribution in Fig. \ref{fig:c}(b) is obtained with an initial probability distribution of the vacuum state, which mainly distributes away from the singular point and thus results in a localized probability distribution along the limit cycle.

{\it Liouvillian spectra.}---To provide further insight into the non-equilibrium dynamics of the quantum system, we also compute the full Liouvillian spectra for three different parameters, corresponding to the time crystal phase [Fig. \ref{fig:d}(a)], equilibrium phase [Fig. \ref{fig:d}(c)] and the critical phase between them [Fig. \ref{fig:d}(b)]. There is only one eigenvalue that is absolutely zero, corresponding to the final equilibrium state.

From the spectra, we can find a frequency scale equal to the detuning $\Delta$. The frequency scale is more clear for smaller driving strengths. Especially, when there is no driving, i.e., $\varepsilon = 0$, the elements in different diagonal lines of the density matrix are independent.
In this case, the Liouvillian operator can be block diagonalized, and the imaginary part of the eigenvalues can be directly obtained as $-im\Delta$. It accounts for the equal spacing spectrum along the imaginary axis in Fig. \ref{fig:d}(a). This property gradually disappears when we increase the driving strength. However, the right several eigenvalues with small real parts preserve the equal spacing frequency \cite{sup}. It is also these eigenvalues that govern the non-equilibrium dynamics and thus the oscillating frequency remains similar to the detuning [see Fig. \ref{fig:d}(d)].

\begin{figure}[b]
	\centering
	\includegraphics[width=0.48\textwidth]{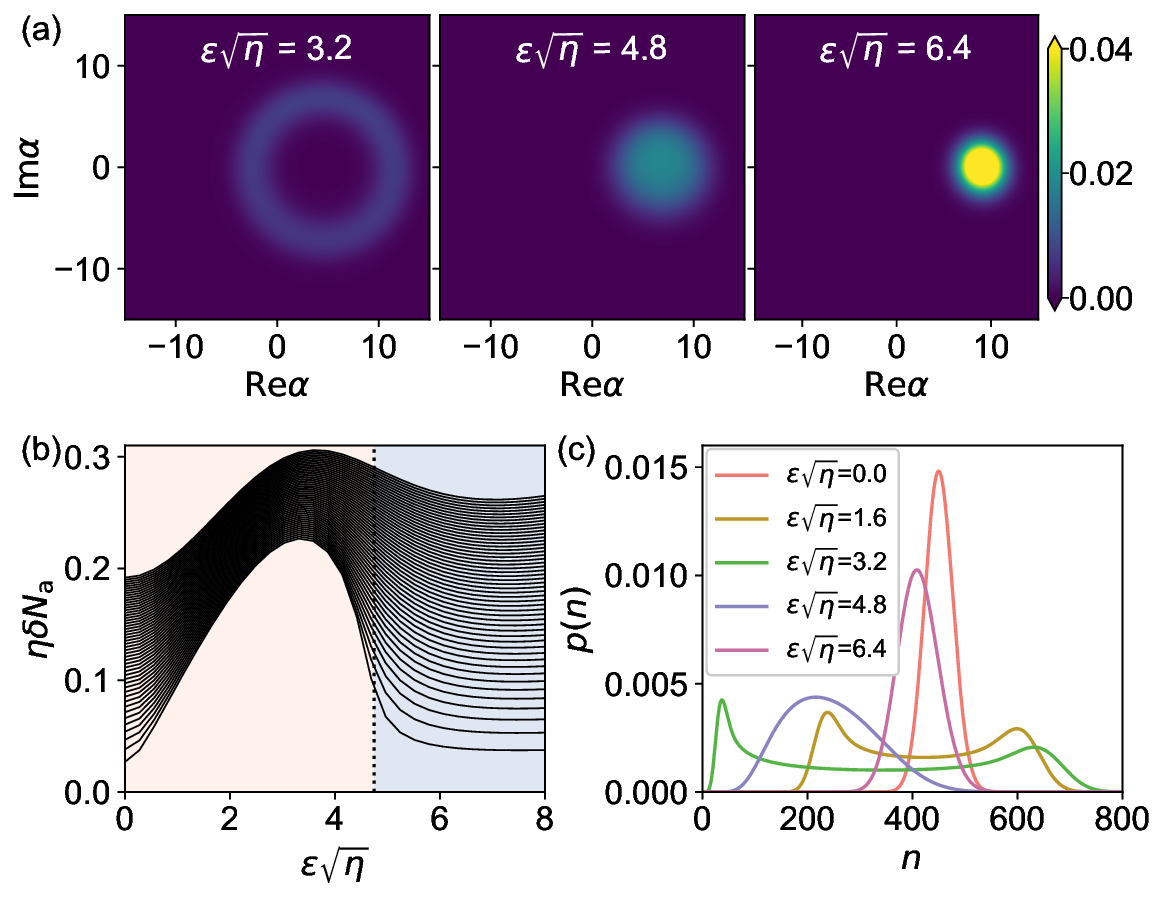}
	\caption{Quantum limit cycle and photon statistics in the steady state. (a) Steady-state quantum Husimi distributions for $\varepsilon\sqrt{\eta}=3.2,4.8,6.4$ from left to right. The nonlinear damping rate is $\eta = 0.01$. (b) The rescaled photon number fluctuation as a function of rescaled driving strength $\varepsilon\sqrt{\eta}$. From the bottom up, the nonlinear damping rate $\eta$ is increased from 0.001 to 0.05 linearly. (c) The photon distributions for $\eta=0.001$ and five different driving strengths. Other parameters are $\Delta = 10$, $\kappa = 0.1$, and $g = 1$.}
	\label{fig:e}
\end{figure}

{\it Dissipative phase transition.}---A dissipative phase transition can be verified from two different aspects. One is the closing dissipative gap as discussed in Fig. \ref{fig:b}(b), and the other is sharp changes of the steady state \cite{PhysRevA.86.012116}. First, the sharp change of the steady state can be seen in the quantum Husimi distribution. As shown in Fig. \ref{fig:e}(a), the steady state has a ring-shaped distribution in the time crystal phase and a Gaussian distribution in the equilibrium phase.

Furthermore, we show in Fig. \ref{fig:e}(b) a steady-state observable, i.e., the photon number fluctuation as a function of the rescaled driving strength. We also observe a sharp transition close to the phase boundary (Hopf bifurcation) in the thermodynamic limit. In the equilibrium phase (and also for $\varepsilon\sqrt{\eta}=0$), the photon number fluctuation decreases to zero exceptionally approaching the classical limit while in the time crystal phase the photon number fluctuation approaches a nonzero value. Consequently, the photon number fluctuation is a good order parameter to characterize the dissipative phase transition. Moreover, as shown in Fig. \ref{fig:e}(c), the nonzero photon number fluctuation in the time crystal phase originates from a broad photon distribution.

{\it Discussion and conclusion.}---The quantum VdP oscillator subject to an external drive has been investigated in detail in several previous works \cite{lee_quantum_2013,walter_quantum_2014,navarrete-benlloch_general_2017,sonar_squeezing_2018}, where they focus more on the quantum synchronization regime instead of the limit cycle regime. A general linearized theory has been developed to describe the quantum fluctuations around the limit cycles \cite{navarrete-benlloch_general_2017}. The long-time dissipative behavior towards the steady state in our model agrees with the prediction of the linearized theory. However, the linearized theory also neglects many details, such as the metastable regime with quantum oscillation, which is a major regime to investigate the time crystal dynamics. Our work provides a full quantum description of the limit cycle phase emerging in the quantum VdP oscillator, and it will deepen our understanding of both the nonlinear quantum oscillator and the time crystal.

The superconducting cavity \cite{leghtas_confining_2015,gertler_experimental_2023} is a promising system to implement this model experimentally. The nonlinear damping can be realized through a four-wave process. Additional gain may be introduced by coupling to an auxiliary system such as another cavity or an atom with a blue-detuned driving. The possibility to measure the Wigner function and photon statistics allows for the investigation of both the time crystal dynamics and dissipative phase transition. The system of trapped irons is another possible experimental platform, as discussed in similar models for the detection of quantum synchronization \cite{lee_quantum_2013,walter_quantum_2014}.

In conclusion, we investigate the non-equilibrium and equilibrium behaviors of the time crystal in a single-mode nonlinear cavity. First, we discuss the well-known classical description of the model. There are two regimes, the limit cycle and the fixed-point regimes, separated by a Hopf bifurcation. Then we show this model in the limit cycle regime possesses pure imaginary eigenvalues of the Liouvillian spectrum in the thermodynamic limit. The imaginary eigenvalues indicate the time crystal dynamics and represent the oscillating period. Moreover, we show after a quick dephasing process the system will enter a metastable regime with the emergence of quantum oscillation. Afterward, the phase fluctuation of the quantum oscillation smears out the limit cycle with a time scale much smaller than the oscillating period. Differently, we find that the oscillating phases of classical phase trajectories are encoded in the initial distribution without a dephasing process. Finally, we show the Hopf bifurcation is related to a dissipative phase transition, and the dissipative phase transition can be characterized by the photon number fluctuation in the steady state. Our results provide a new way to investigate the time crystal from the dissipative phase transition and may find application in quantum associative memory.

\begin{acknowledgments}
	This work is supported by the National Key R\&D Program of China (Grant No. 2023YFA1407600), and the National Natural Science Foundation of China (NSFC) (Grants No. 12275145, No. 92050110, No. 91736106, No. 11674390, and No. 91836302).
\end{acknowledgments}

%

\end{document}